\documentclass[prd,twocolumn,nofootinbib]{revtex4}
\usepackage{latexsym}
\usepackage{amsthm}
\usepackage{ulem}
\usepackage{amsmath}
\usepackage{graphicx}  % needed for figures
\usepackage{bm}        % for math
\usepackage{amssymb}   % for math
\usepackage{hyperref}
\usepackage{color}
\begin{document}
\title{Causal loop in the theory of relative locality}
\author{Lin-Qing Chen}
\affiliation{Perimeter Institute for Theoretical Physics, Waterloo, Ontario N2L 2Y5, Canada}
\begin{abstract} 
We find that relative locality, a recently proposed Planck-scale deformation of special relativity, suffers from the existence of causal loops. A simple and general construction of such on-shell loop processes is studied. We then show that even in one of the weakest deformations of the Poincar\'e group in relative locality, causality can be violated. 
\end{abstract}
\maketitle
The search for quantum gravity has led to the idea that special relativity should be modified at high energies in such a way that (modified) Lorentz transformations leave the Planck scale invariant. Doubly special relativity (DSR) has been proposed as an embodiment of this idea \cite{AmelinoCamelia:2000mn, Magueijo:2001cr}. However, it was shown in \cite{Hossenfelder:2010tm} that this theory has observer dependent non-localities for distant interactions. Relative locality is a reincarnation of DSR that tries to clarify the issue by proposing a radically different way of thinking about physics.

The birth of relative locality came from the insight that we never directly observe the spacetime we postulate. Our usual picture of spacetime is constructed operationally from the measurements of energies, momenta and times of events\cite{AmelinoCamelia:2011bm, AmelinoCamelia:2011pe}. The notions that everything shares a universal spacetime and that locality has absolute observer-independent meaning  could be  unwarranted assumptions. 
As a theory describing quantum gravity induced modifications to relativistic dynamics of particles, relative locality was proposed at the regime where $\hbar \rightarrow 0, G  \rightarrow 0$, while their ratio $m_{P}= \sqrt{\hbar/G}$ is held fixed for every observer\cite{AmelinoCamelia:2011bm, AmelinoCamelia:2011pe}. So effects due to the presence of the Planck mass are expected. Some RL phenomology has been studied in \cite{AmelinoCamelia:2011bm, AmelinoCamelia:2011pe, Freidel:2011mt, AmelinoCamelia:2011nt}

Relative locality takes momentum space $\mathcal{P}$ as primary and formulates classical dynamics on the phase space $T^*(\mathcal{P})$\cite{AmelinoCamelia:2011bm}. The geometry of momentum space is not preassumed to be that of a linear space but could have curvature, torsion and non-metricity in general, and should be tested by experiments. There is no global projection that gives a description of processes in a universal spacetime. 
The notion of absolute locality and universal spacetime is equivalent with the assumption that
 the conservation law of momenta is linear. \cite{AmelinoCamelia:2011bm}. 

We start with giving a brief review of the classical dynamics and phase space structure of relative locality. After defining the causal structure in RL, we go on to show that the theory has solutions that are causal loops. The general conditions allowing for such loops to happen are then studied. We illustrate this construction when the geometry of momentum space is taken to be that of $\kappa-$Poincar\'e, which is the most well studied and the simplest non-trivial geometry of momentum space in RL \cite{AmelinoCamelia:2011nt, Gubitosi:2011ej}. The appearance of the causal loops is a result of the phase space structure of the theory and  the causal loops vanish in the limit of special relativity.The existence of causal loops implies that it would be non-trivial to construct a quantum theory of RL with unitarity, a worry which should be addressed in future research. If RL were shown to be a well-tested theory in the future, causal loops will challenge our understanding of the fundamental role of causality. More probably, this result implies that RL is incomplete or wrong.

\vspace{-3mm}
\section{Classical particle dynamics and phase space in relative locality}
\vspace{-3mm}
Momentum space is assumed to be a manifold $\mathcal{P}$ with a metric $g^{ab}$ and a connection $\Gamma^{ab}_c$. The geodesic distance $\mathcal D(p)$ from the origin to a point $p\in\mathcal{P}$ is interpreted as the rest mass of a particle with momentum $p$. The non-linear addition rule of momenta $p\oplus q$, which should be found experimentally, determines the connection: 
\vspace{-1 mm}
\begin{equation}
\frac{\partial}{\partial p_a} \frac{\partial}{\partial q_b} (p \oplus_k q)_c\big|_{p=q=k}=-\Gamma^{ab}_c(k) 
\vspace{-1mm}
\end{equation}
Introduce an inverse operation $\ominus$ satisfying $(\ominus p)\oplus p=0$ which turns incoming momenta into outgoing momenta. Now we can write the conservation law associated with any interaction vertex as a non-linear equation $\mathcal {K}_a(p^I) \equiv 0$. For example, a three vertex can be written as $(p \oplus q) \ominus k=0 $. The order of addition is important, since it corresponds to micro-causal structure of interactions\cite{AmelinoCamelia:2011bm}.
The torsion of momentum space is a measure of the non-commutativity and the curvature is a measure of non-associativity of addition rule of momenta.\cite{AmelinoCamelia:2011bm, Freidel:2011mt, Gubitosi:2011ej} 

Dynamics of point particles is defined by the action:
\vspace{-1 mm}
\begin{equation}
\begin{split}
\!\!S&=\sum_J{S^J_{free}}+\sum_i S^i_{int}\\
&=\sum_J \! \int{\!ds(x_J^a \dot{p}^J_a +\! \mathcal N_J \mathcal C^J (p^J))}  + \sum_i  \mathcal K^i_a(p^J(s_i)) z_i^a
\end{split}
\vspace{-1mm}
\end{equation}
where $s$ is an affine (time) parameter along the trajectory of the particle and an interaction labeled by $i$ happens at $s_i$ for each particle; $x_J$ are \textit{Hamiltonian spacetime coordinates} which are defined as being canonically conjugate to $p^J_a$:  $\{x_I^a,p^J_b\} = \delta^a_b \delta^J_I$ and $x_J^a \in T^*_{p^J}$. The mass shell condition
$\mathcal C^J(p)\equiv \mathcal D^2(p^J)-m^2_J $ is  imposed by the Lagrange multiplier $\mathcal N_J$. The interaction part of the action is a Lagrange multiplier times the conservation of momenta 
$\mathcal K_a (p^1, p^2...)\equiv 0$. 
By varying the action, we get the equations of motion, the two of which we will be concerned with:
\vspace{-1 mm}
\begin{eqnarray}
u^a_J \equiv\dot{x}^a_J =\mathcal N_J \frac{\partial \mathcal C(k^J)}{\partial k_a^J}
\label{four velocity} \\
x^a_J(s_i)=\pm z^b_i \frac{\partial \mathcal K^i_b}{\partial k^J_a}
\label{end points}
\vspace{-1 mm}
\end{eqnarray}
$\pm$ indicates an incoming/outgoing particle respectively.
Equation (\ref{four velocity}) tells us how free particles propagate on one cotangent space (``Hamiltonian spacetime" $T^*_p$) of their phase space; and equation (\ref{end points}) describes how cotangent spaces $T^*_p$ of different particles are connected by interaction events $z$. 

Physics should be invariant under momentum space diffeomorphisms, i.e. redefinition of coordinates on the momentum manifold. Such a  transformation is given by: $p_a\! \rightarrow\! \tilde{p}_a=f_a(p)$ s.t. the geodesic distance between the origin and $p$ is preserved. Under this transformation, $\mathcal{K}(p)\! \rightarrow\! \tilde{\mathcal{K}}=\mathcal{K} (f(p))$,  $x^a$ transforms like a covector: $x^a\! \rightarrow \! \tilde{x}^a=x^b\left(\partial{f_a(p)}/\partial{p_b} \right)^{-1} \in T^*_p, z^a \! \rightarrow\! \tilde{z}^a= z^c (\delta{\tilde{\mathcal{K}}_a}/\delta{\mathcal{K}_c} )^{-1}$. By applying these transformations to our action we see that it is unaltered, as desired.

\vspace{-3mm}
\section{Causal loop process}
\vspace{-3mm}

In relativity the causal relationships between events are expressed as geometrical relationships between points on the spacetime manifold, while in RL we do not have a universal spacetime. We thus have to come back to the most fundamental notion of causality between events. Define event B to be in the {\textit{causal past}} of event A if  in the process that is being considered, there exists a sequence of events $B, B_1, ... B_n, A, n\geq 0$ s.t. from each event there exists an outgoing free-propagating particle coming to the next event. We write it as $B\!\prec \!A$. Vice versa we can define {\textit{causal future}} $C\!\succ \!A$. The above definition of causal relationship is actually a strict partial order. In analogy with the notion in general relativity, one type of causality violation is if $\exists$ events $A, B$ s.t.$A\!\prec \!B, B\!\prec \!A$ i.e. causal loop. 

\begin{figure}[h]
  \centering
    \includegraphics[width=0.36\textwidth]{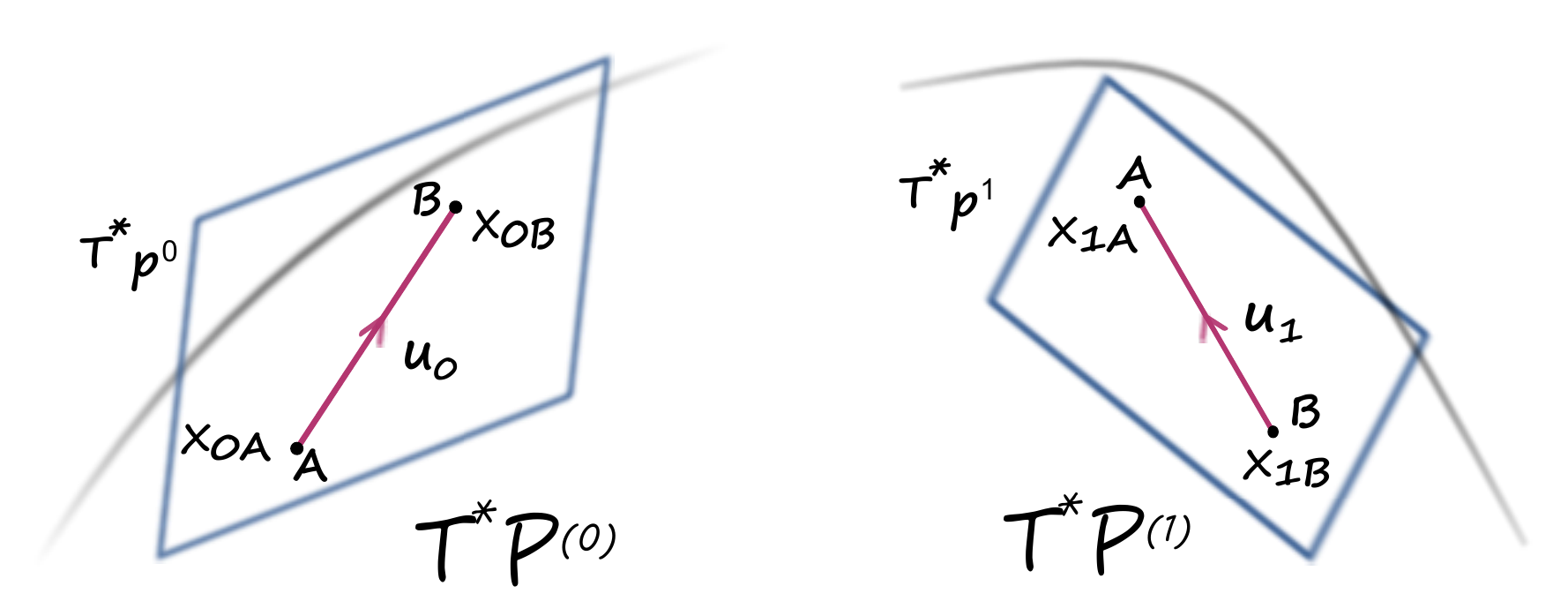}
\vspace{-1.5 mm}
	\caption{A causal loop on the two-particle phase space.}
\label{phasespace}
\vspace{-3mm}
\end{figure}

Let us check the existence of the simplest causal loop.  Assume two bi-particle collision events $A$ and $B$, defined by conservation laws $\mathcal{K}^A \! \equiv 0$ and $\mathcal{K}^B \!\equiv 0$. Particle 0 with momentum $p^0$ is created from event $A$ and then collides with another particle at event $B$.  The twist is now to consider particle 1 with momentum $p^1$ created at event $B $ and then colliding with another particle at event $A$, which creates the particle 0. We use $x_{0A}$ and $x_{0B}$ to label the starting point and ending point of particle  $p^0$'s worldline which lives on $T^*_{p_0}$; and similarly $x_{1B}$ and $x_{1A}$ for particle $p^1$, as in Fig[\ref{phasespace}]. 

Particle 0 freely proporgates from event A to B for proper time $\tau_0$ on $T^*_{p^0}$; the end of its worldline is related with the  starting point of  particle 1's worldline by interaction B:
\vspace{-1 mm}
\begin{equation}
 \left( x^\mu_{0A}+u^\mu_0 \tau_0 \right) \left(\frac{\partial{\mathcal{K}^B_{\alpha}}}{\partial p^0_{\mu}}\right)^{-1}   =x_{1B}^\nu\left(\!-\frac{\partial{\mathcal{K}^B_{\alpha}}}{\partial p^1_{\nu}}\right)^{-1} 
\vspace{-1mm}
\end{equation}
Similarly, the other part of the loop is described by
\vspace{-1 mm}
\begin{equation}
\left( x_{1B}^\nu+u^\nu_1 \tau_1 \right) \left(\frac{\partial{\mathcal{K}^A_{\alpha}}}{\partial p^1_{\nu}}\right)^{-1} =x_{0A}^\mu \left(\!-\frac{\partial{\mathcal{K}^A_{\alpha}}}{\partial p^0_{\mu}}\right)^{-1} 
\vspace{-1mm}
\end{equation}
where $u_0, u_1$ are given by equation of motion (\ref{four velocity}).

Thus the existence of a causal loop process is equivalent to the following equation having a physical solution for $x^\mu_{0A}$ and proper times $\tau_0, \tau_1$:
\vspace{-1 mm}
\begin{equation}
\!\! \left(\mathcal{M}_A\!-\!\mathcal{M}_B \right)_\mu^\nu x_{0A}^{\mu}\!=\tau_0\left(\mathcal{M}_B \right)_\mu^\nu\! u^{\mu}_0\!+\!\tau_1 u_1^{\nu}, \ \tau_0, \tau_1\!\in\!\mathbb {R}_+\!\!
\label {general condition}
\vspace{-1 mm}
\end{equation}
where the matrix $\left(\mathcal{M}_A\right)_\mu^\nu := \left(\partial{\mathcal{K}^A_{\alpha}}/\partial p^1_{\nu}\right) \cdot \left(-\partial{\mathcal{K}^A_{\alpha}}/\partial p^0_{\mu}\right)^{-1} $ and $\left(\mathcal{M}_B\right)_\mu^\nu :=\left(-\partial{\mathcal{K}^B_{\alpha}}/\partial p^1_{\nu}\right) \cdot \left(\partial{\mathcal{K}^B_{\alpha}}/\partial p^0_{\mu}\right)^{-1}$.
\\

The above is a general condition. Once we specify the geometry of momentum space and write down the conservation laws of two vertices, the above condition (\ref {general condition}) will then give a system of four linear equations with six free unknowns: $x^\mu_{0A}$, $\tau_0$ and $\tau_1$.

Based on the Cramer's rule, if the matrix $\left( \mathcal{M}_A\!-\!\mathcal{M}_B \right)$ has rank four, i.e. the determinant is non-zero, 
any physical choice of $\tau_0$ and $\tau_1$ yields a unique solution for $x^\mu_{0A}$. If the matrix $\left(\mathcal{M}_A\!-\!\mathcal{M}_B \right)$ has rank less than four, we can fix a ratio $\tau_1=\beta \tau_2, \beta \in \mathbb {R}_+$ first and then check the  number of  remaining unknowns of the system of the  homogeneous equations. 
Assuming the number of unknowns is $d$ ($d \leq 5$ now), if $\exists  \beta$ s.t. the new matrix of coefficients of the system of equations has rank $< d$, then we can always get at least one  physical solution for $\tau_1 >0$. When for $\forall \beta \in\mathbb {R}_+$  the new  coefficients matrix has a rank $\geq d$, only then the equation does not have physical solutions, which means the causal loop (corresponding to these specific interaction vertices) does not occur.

The solution will be invariant under the momentum space diffeomorphsim, because the equation (\ref{general condition}) is just based on the equations of motion, which are invariant.

In the special relativity limit of RL, beacuse of the linearity of momentum space, which also implies trivial isomorphism between cotangent spaces, the transport operators  $\mathcal{M}_A$, $\mathcal{M}_B$ are all identity matrices. This makes the above equation (\ref {general condition}) immediately degenerate to  $\tau_0 u_0^a+\tau_1 u_1^a=0$, which doesn't have physical solution. It means that this kind of causal loop will not occur in special relativity.

The causal loop formed by two interaction events is just the simplest case. Similar processes can be constructed by other forms of vertices and more events: $A\!\prec \!B$,  $B\prec...N$, $N\!\prec \!A$, which conditions enjoy the similar form with the simplest case:
\vspace{-1 mm}
\begin{equation}
\begin{split}
&\left(\mathcal{M}_{A}\!-\!\mathcal{M}_B \mathcal{M}_C...\mathcal{M}_n \right)_\mu^\nu  x_{1A}^{\mu}\!=\! \tau_1  \mathcal{M}_B ...\mathcal{M}_n  u_1^{\nu}+\\
&+\tau_2 \mathcal{M}_C... \mathcal{M}_n u_2^{\nu}+ ...+ \tau_{n-1} \mathcal{M}_n u_{n-1}^{\nu}+ \tau_n u_n^{\nu}
\end{split}
\vspace{-1mm}
\end{equation}
These compose a system of four linear equations with $4+n$ unknowns, where $n$ is the number of interactions that form the causal loop. Compared with eq. (\ref {general condition}), the above conditions are even easier to be satisfied, since there are more unknowns in the same number of equations.  More general causal loops that contain branching out can be always decomposed into a few simple loops described by the above process. 

\vspace{-3 mm}
\section{Two-event causal loop in $\kappa$-Poincar\'e momentum space}
\vspace{-3mm}
In this section we will illustrate the existence of causal loops in a specific geometry of momentum space. $\kappa$-Poincar\'e Hopf algebra, a dimensionful deformation of the Poincar\'e group, describes a momentum space with de Sitter metric, torsion and nonmetricity, which is the first well-studied example of the non-trivial geometry of momentum space in relative locality  \cite{Majid:1994cy},  \cite{Gubitosi:2011ej}, \cite{AmelinoCamelia:2011nt}.  It is also one of the weakest deformation of Poincar\'e group as a Hopf algebra\cite{Majid:1994cy}.
The line element of the $\kappa -$Poincar\'e momentum space in comoving coordinates is given by:
\vspace{-1mm}
\begin{equation}
ds^2=dp_0^2 - e^{2p_0/\kappa} \delta^{ij} dp_i dp_j \ \ \ i,j=1,2,3 \!\!\!
\vspace{-1.5 mm}
\end{equation}
where $\kappa$ is a large energy scale close to Planck energy. The mass-shell condition is given by the geodesic distance from point $p$ to the origin of momentum space\cite{Gubitosi:2011ej}:
\vspace{-0.5 mm}
\begin{equation}
m(p)= \kappa Arccosh(\cosh (p_0/\kappa)-e^{p_0/\kappa} |\vec{p}|^2 /2\kappa^2)
\vspace{-1 mm}
\end{equation}
from which we can get e.o.m.(\ref{four velocity}). The addition rule of momenta on $\kappa -$Poincar\'e momentum space is as follows\cite{Gubitosi:2011ej}:
\vspace{-5.5 mm}
\begin{equation}
(p\oplus q)_0 =p_0+q_0\ \ \ 
(p\oplus q)_i =p_i+ e^{-p_0/\kappa} q_i \!\!
\vspace{-1 mm}
\end{equation}

\begin{figure}[h]
  \centering
    \includegraphics[width=0.18\textwidth]{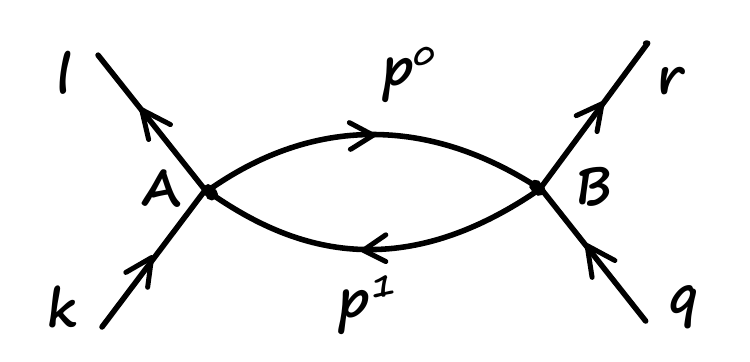}
\vspace{-1.5 mm}
	\caption{A two-event causal loop in relative locality.}
\label{loop}
\vspace{-3mm}
\end{figure}

As an example of the simplest causal loop,  we look at two events $A B$, which non-linear conservation of momenta are $\mathcal{K}_A \! =( k \oplus p^1) \ominus (p^0 \oplus l)\equiv 0, \ \mathcal{K}_B\!=(p^0 \oplus q) \ominus (r \oplus p^1)\equiv 0$, see fig. \ref{loop} .

\vspace{-0mm}
%\noindent

The conservation of momenta can be writen out explicitly  as (\ref{addition rule}).  
\vspace{-1 mm}
\begin{equation}
\begin{split}
(\mathcal{K}^A)_0&=( k_0 + p^1_0) - (p^0_0 + l_0)\\
(\mathcal{K}^A)_i&=k_i+p^1_i e^{-k_0/\kappa} - e^{\frac{1}{\kappa} (-\mathcal{K}^A_0)} \left(p^0_i  + l_i e^{-\frac{1}{\kappa}p^0_0}\right)\\
(\mathcal{K}^B)_0&=(p^0_0+q_0)-(r_0+p^1_0)\\
(\mathcal{K}^B)_i&=p^0_i+q_i e^{-p^0_0/\kappa}-e^{\frac{1}{\kappa}(-\mathcal{K}^B_0)}\left(r_i +p^1_i e^{-\frac{1}{\kappa}r_0}\right)
\end{split}
\label{addition rule}
\vspace{-2 mm}
\end{equation}

Now, we have all the necessary elements to calculate the condition (\ref{general condition}) for this specific example. It turns out that the matrix $\left( \mathcal{M}_A\!-\!\mathcal{M}_B \right)$ has rank three and an unknown $x^0_{0A}$ drops out, which is due to the linear addition rule of the zero component of momenta, in (\ref{addition rule}). For simplicity, assume that the particles with momentum $p^0, p^1$ have same rest mass $m$. Using $\tau_0$ to rescale the other unknowns and solve the above equation for $\tau_1/\tau_0$ and $x^i/\tau_0, i=1,2,3$, we get the solution, in which the ratio of two proper times is
\vspace{-0.8 mm}
\begin{equation}
\!\!\!\frac{\tau_1}{\tau_0}\!=\frac{p^0_i e^{\frac{p^0_0}{\kappa}}\!\left[F p^0_i \!+ 2r_i e^{\frac{r_0}{\kappa} }\! -2k_i e^{\frac{k_0}{\kappa} } \right]\!\!+\!2F \kappa^2 \sinh (\frac{p^0_0}{\kappa})}{p^1_i e^{\frac{p^1_0}{\kappa}}\! \left[2 e^{\frac{k_0+r_0}{\kappa}} (k_i\!-\!r_i) -\!F p^1_i\right]\!\!-\!2F \kappa^2 \sinh (\frac{p^1_0}{\kappa })}
\label{solution}
\vspace{-1 mm}
\end{equation}
where $F=e^{k_0/\kappa}-e^{r_0/\kappa}$ just for shorthand. Thus as long as there exists physical values of momenta such that the solution for $x_{0A}^i/\tau_0$ is finite and  $\tau_1/\tau_0$ is positive, this specific causal loop can occur. We check it as follows.

\begin{figure}[h]
  \centering
    \includegraphics[width=0.31 \textwidth]{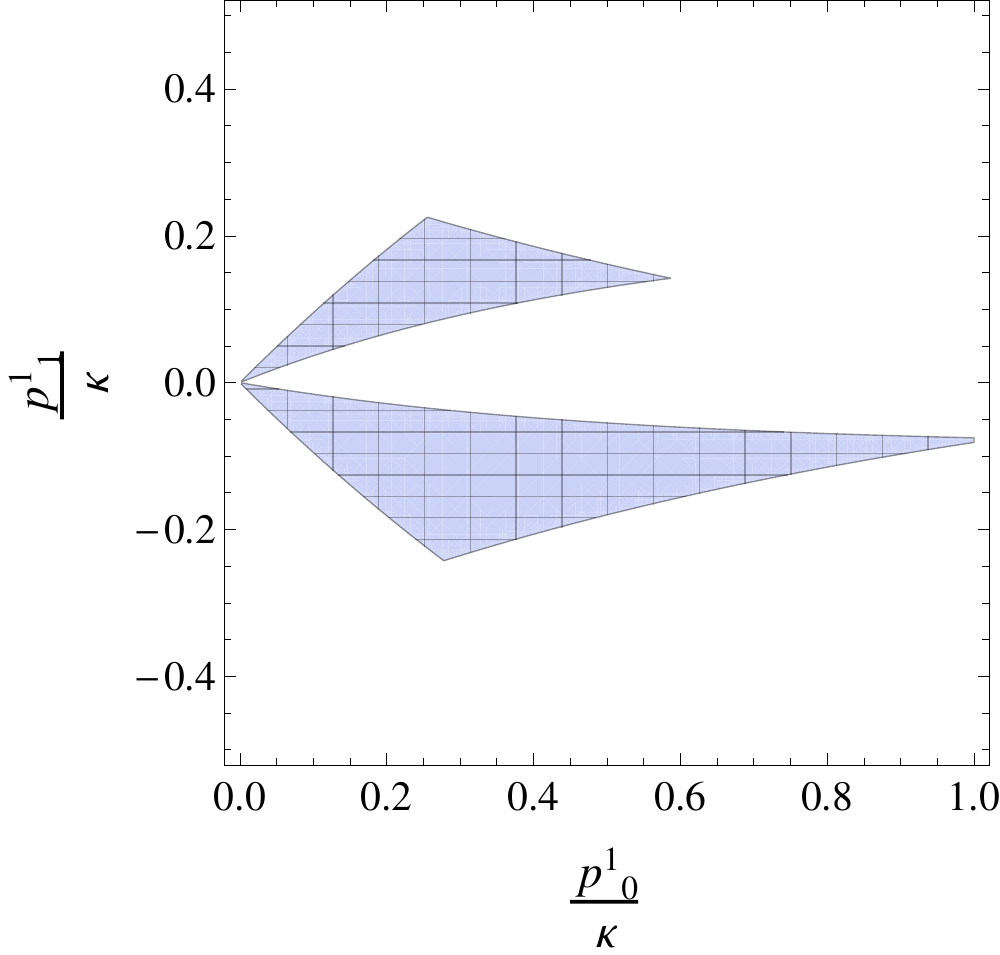}
\vspace{-3 mm}
	\caption{ The region of $\left( p^1_0, p^1_1\right)$ that possiblly leads to a causal loop by events A \& B when $p^0_0=p^1_0, p^0_1=-p^1_1, r_0= 1.1 k_0 = 0.88\kappa, k_1=-r_1=0.1 \kappa $. }
\label{region}
\vspace{-3.5 mm}
\end{figure}

For simplicity and without lose of generality, we set the last two spatial components for all the momenta to be zero.  Because the energy of single particle has to be smaller than $\kappa$, energies $p^0_0,p^1_0,k_0,r_0 \in (0,\kappa)$. The requirement of timelike on-shell momenta  $m(p) >0$ constrains the spatial components, 
\vspace{-1mm}
\begin{equation}
e^{-p_0/\kappa} -1< p_1/\kappa < -e^{-p_0/\kappa} +1, \ \ \  p_2, p_3=0
\vspace{-1 mm}
\end{equation}
We can then plot the region in terms of $\left( p^1_0, p^1_1\right)$ that allow the equation to have physical solution for $x$ and $\tau_1, \tau_2$ under a specific relationship among other momenta. See Fig.\ref{region}

\vspace{-2 mm}

Surprisingly, for any low energy scale of momenta choice (as long as $\kappa$ is not infinity), there always exist solutions for positive $\tau$s and finite $x_{0A}$.  At low energy, the solution of $x_{oA}$ in eq. (\ref{general condition}) is proportional to $\kappa/m$:

\vspace{-2 mm}
\begin{equation}
x^1_{0A} \approx \frac{ (p^1_0 p^0_1- p^0_0 p^1_1) \tau_0 \kappa}{[p^1_0(k_0-r_0)+p^1_1 (r_1-k_1) ]m}
\vspace{-0.5 mm}
\end{equation}
thus when the scale $\kappa$ is much larger than the scale of momenta, $x_{0A}$ has quite large scale compared with $\tau_0$. At high energy, the scales of solutions are reasonable. Two random examples taken from two different energy scales are shown in Table (\ref{tab}). If the current form of relative locality were to be correct, then we would expect causal loops to be common at high energy.

\vspace{-6mm}
\begin{table}[h]
\caption{Comparison of the results at different energy scales} 
\centering % centering table
\begin{tabular}{c c c} % creating 10 columns
\hline\hline % inserting double-line
   & High energy scale &Low energy scale 
\\ [0.5ex]
\hline % inserts single-line
% Entering 1st row
$p^0/\kappa$& $(0.5, 0.1,0,0)$ &$(10^{-22}, -0.2\times 10^{-22},0,0)$\\
$p^1/\kappa$&$ (0.5, -0.1,0,0) $&$(10^{-22}, 0.2\times 10^{-22},0,0)$\\
$k/\kappa$&$ (0.8, 0.1,0,0) $&$(10^{-22}, 10^{-23},0,0)$\\
$r/\kappa$&$ (0.85, -0.1,0,0) $&$(1.01 \times 10^{-22},-10^{-23},0,0)$\\
%$l/\kappa$&$ (0.8, -0.07,0,0) $&$(10^{-22},5\times 10^{-23},0,0)$\\
%$q/\kappa$&$ (0.85, -0.4,0,0) $&$(1.01 \times 10^{-22},3\times 10^{-23},0,0)$\\
$m/\kappa$ &$0.48  $&$ 9.80\times10^{-23}$\\
$\tau_1/\tau_0$&$ 1.2 $&$0.6$\\
$x^\mu_{0A}/\tau_0$&$(\forall x^0_{0A} \in \mathbb{R}, 11.8, 0,0)$&$(\forall x^0_{0A} \in \mathbb{R}, 8.2 \times 10^{22}, 0,0)$\\
[1ex]
% [1ex] adds vertical space
\hline % inserts single-line
\end{tabular}
\label{tab}
\vspace{-2mm}
\end{table}
\vspace{-6 mm}
\section{Discussion and Conclusions}
\vspace{-4 mm}
We have shown that  relative locality allows causal loops solutions, and it is surprisingly simple to construct them. This is a generic feature arising from the phase space structure of RL: there is no universal spacetime, and the configuration information lives on the different cotangent spaces of the momentum manifold. The non-linear conservations of momenta at interaction events determine how the cotangent spaces $T^*_p$ of different particles are connected by the interaction, Eq.(\ref{end points}). Due to the nontrivial relations between cotangent spaces, a particle can come back to the event that causes its own creation as one of the incoming participants of that interaction. 

In the last section, the causal loop's existence depends on specific choices of points on the cotangent space $T^*_p$ (let us call it``x dependence"). 
In general, after fixing the proper times, the solution set of $x^\mu_{oA}$ (if it exists) is a lower dimensional subspace. In the example we calculated, it was a line on $T^*_{p^0}$.  The same unusual feature of ``x dependence" is also present in many of the usual loop processes without causal issues in relative locality, e.g.\cite{CamoesdeOliveira:2011gy}.  
Some nets of vertices lead to the dependence on ``where'' the events are on the cotangent space, while some do not. We do not yet have fundamental reasons for choosing some forms of vertices rather than others just for the sake of ``x independence". 
This necessitates future work in understanding the physical meaning of the Hamiltonian spacetimes and the choices of vertices. 

In \cite{AmelinoCamelia:2011nt} authors enforced the translation invariance on $T^*_p$, which is a more strict symmetry than ``x independence". However, the approach there requires the use of very non-local interaction vertices to achieve the symmetry, which is not physical. Future work should address the following questions. Is ``x dependence" of many loops a generic feature? Should translation invariance on $T^*_p$ be a fundamental symmetry of relative locality? How do we achieve  ``x independence" or translation invariance on $T^*_p$  in a physical way? Can we remove the causal loops by enforcing those symmetries in the theory?

Does the existence of causal loops imply that there is an inconsistency in the theory?  In general relativity, Einstein equations have closed-timelike-curve (CTC) solutions.  One way out of the grandfather paradox is the Novikov self-consistency principle, which states that events in CTC influence each other in a self-adjusted, cyclical way and the only solutions can occur locally are those which are globally self-consistent \cite{Friedman:1990sx}. However, it has been shown that at the quantum level, unitarity fails for interacting fields in CTC and the subjective probabilities of events can be different for different observers \cite{Friedman:1992jc, Deutsch:1991nm, Boulware:1992pm}. It probably points towards that spacetimes with CTC are unphysical. In relative locality, a well-defined quantum theory has not been built yet, but there is ongoing research into constructing it. The lesson from quantum field theory in curved spacetime shows that it would be non-trivial to have causal loops in a consistent unitary theory. It is thus essential to study whether it is even possible to construct a unitary quantum field theory for relative locality.

If the current form of relative locality were to be experimentally established in the future, we need to rethink causality as a fundamental property of nature. It would influence some approaches to quantum gravity which take discrete causal structure as basic assumption. Another possibility is that relative locality is an incomplete or a wrong theory in its current form, and the revision of it should exclude causal loops.
\noindent

{ \bf Acknowledgements:}
I am very grateful to L. Smolin, L. Freidel, G. Amelino-Camelia,  A. Banburski,  T. Rempel, S. Hossenfelder, J. Kowalski-Glikman, F. Mercati,  J. Hnybida, T. Wang and J. Ziprick for helpful discussions and suggestions. 
Research at Perimeter Institute is supported by the Government of Canada through Industry Canada and by the Province of Ontario through the Ministry of Economic Development \& Innovation. The work was supported by NSERC and FQXi.
\vspace{-4 mm}
%%%%%%%%%%%%%%%%%%%%%%%%%%%%%

\end{document}